\documentclass[manuscript]{aastex}
\usepackage{amsmath}
\DeclareMathOperator{\sech}{sech}
\usepackage{epstopdf}

\usepackage[utf8]{inputenc}
\DeclareUnicodeCharacter{FB01}{fi}
\DeclareUnicodeCharacter{2212}{-}
\usepackage{natbib}

\usepackage{graphicx}

\begin{document}


\title {Gravitational microlensing constraints on primordial black holes by Euclid}


\author{Lindita Hamolli\altaffilmark{}}\and\author{Mimoza Hafizi \altaffilmark{}}
\affil{Department of Physics, University of Tirana, Albania.}
\email{lindita.hamolli@fshn.edu.al}
\and
\author{Francesco De Paolis\altaffilmark{}}\and\author{Achille A. Nucita \altaffilmark{}}
\affil{Department of Mathematics and Physics "Ennio De Giorgi", University of Salento, Via Arnesano, 73100 Lecce, Italy}
\affil{INFN, Sezione di Lecce, Via Arnesano, 73100 Lecce, Italy}


\begin{abstract}
Primordial black holes (PBHs) may form in the early stages of the Universe via the collapse of large density perturbations. Depending on the formation mechanism, PBHs may exist and populate today the galactic halos and have masses in a wide range, from about $10^{-14} \, M_{\odot}$ up to thousands, or more, of solar masses.  Gravitational microlensing is the most robust and powerful method to constrain primordial black holes (PBHs), since it does not require that the lensing objects be directly visible. We calculate the optical depth and the rate of microlensing events  caused by  PBHs eventually distributed in the Milky Way halo, towards some selected directions of observation. Then we discuss the capability of Euclid, a space-based telescope which might perform microlensing observations at the end of its nominal mission, to probe the PBH populations in the Galactic  halo.
\end{abstract}


\section{Introduction}
Since their proposal, Primordial Black Holes (PBHs) have been  objects of intensive study (see, e.g., \cite{hawking1974black}).  More recently, the interest of the scientific community towards these objects increased substantially and they are nowadays often considered as a viable dark matter (DM) candidate (\cite{frampton2020primordial}).
The mechanism of PBH formation involves large density  fluctuations or inhomogeneities  in the early universe. Indeed, we know that density fluctuations should be present in the early stages of the Universe in order to seed structure formation. Then, provided that matter/energy is compressed to a high enough density, that is to a size as small as its Schwarzschild radius ($R_S \equiv \frac{2GM}{c^2}$), PBHs may form.  In particular, Carr and Hawking (\citeyear{carr1974black}) have shown that if over-densities of order unity exist in the early universe then a PBH certainly forms when the perturbed region enters cosmological horizon. It is not difficult to show that the PBH mass grows linearly with the formation time $t$  (\cite{carr2020primordial}):

\begin{equation}
    M_{PBH}\simeq \frac{c^3 t}{G} \simeq 10^{15}\left(\frac{t}{10^{-23} \,s}\right)\,g \simeq 10^{5}\left(\frac{t}{1 \,s}\right)M_{\odot}.
    \label{masa}
\end{equation}

Then, depending on the PBH formation epoch, black holes of different masses can form in the early universe. It has been shown  that PBHs with mass smaller than about $10^{15}$ g would have evaporated by now, so attention has shifted to the PBHs with larger mass.
Hawking thought that PBHs would have  formed before the electroweak phase transition, which occurred at  $t\simeq 10^{-12}$ s. In this case one finds $M_{PHB}\leq 10^{-7} \, M_{\odot}$.
However,  there is no reason in principle which forbids the formation of PBHs later.  In this case, the PBH mass can  rise of orders of magnitude, even up to $\sim 10^{5}\, M_{\odot}$ or so. Certainly, only observations might be able to settle this open issue.
 
PBHs might lead to various interesting astrophysical consequences, such as providing seeds for supermassive black holes in galactic nuclei (\cite{bean2002could}, \cite{garcia2017massive}, \cite{dolgov2019massive}), influence the generation of large-scale structure through Poisson fluctuations (\cite{afshordi2003primordial}) and cause important effects on the thermal and ionization history of the Universe (\cite{ricotti2008effect}). But, perhaps, the most exciting possibility is that they could provide a non-negligible fraction  of the dark matter, which is known to constitute about $25\%$ of the critical density of the Universe (\cite{carr2016primordial}).

There are some observational conundra that may be explained by PBHs. They are associated with lensing (\cite{niikura2019constraints}, \cite{mediavilla2017limits}),  accretion (\cite{cappelluti2013cross}), dynamical (\cite{clesse2018detecting}) and  gravitational wave (GW) effects (\cite{abbott2016ligo}).  
Various constraints on their abundance allowed to conclude that PBHs in  three ranges of masses could provide the dark matter (\cite{barrau2004peculiar}): the asteroid mass range ($10^{-17} - 10^{-16}~M_{\odot}$), the sublunar mass range ($ 10^{-13} -10^{-9}~ M_{\odot}$) and what is sometimes called the intermediate-mass black hole (IMBH)  range $(10-10^{3}~ M_{\odot})$.

The IMBH possibility is  of special interest due to the recent detection of black-hole mergers by LIGO/Virgo (\cite{abbott2016ligo}) with masses somewhat larger than those expected from stellar evolution. In general, constraints from these observables have been computed under the assumption that all PBHs have the same mass. The corresponding mass function, comprising a single Dirac delta function, is said to be monochromatic. However, an extended (non-monochromatic) mass function, which will correctly combine constraints across all masses, is necessary.
Recently, considering PBH formation by the collapse of energy density fluctuations with a Gaussian probability
distribution, Sureda et al. (\citeyear{sureda2020press}) derived an extended mass function for PBHs based on a modified Press-Schechter (PS) formalism. They have studied two different PBH formation scenarios named Fixed Conformal Time, where all the PBHs are formed at the same epoch, and Horizon Crossing, where a PBH is formed when the appropriate fluctuation scale reenters the cosmological horizon.

Concerning the fraction $f(M)$ of PBHs in the Galactic halo, the currently available limits are summarized in Fig.1 of Carr $\&$ Kuhnel \citeyear{carr2020primordial}. The main constraints are derived  from PBH evaporation, different gravitational-lensing experiments, numerous dynamical effects and PBH accretion.
Even if many different methods have been proposed in the literature for detecting and constraining PBHs (see, e.g.  (\cite{carr2016primordial},  \cite{green2021primordial}) and references therein) \footnote{In addition to the methods discussed in that paper we also mention that lensing by fast radio bursts (FRBs) can allow constraining PBHs with mass larger than about $2\, M_{\odot}$  (\cite{laha2020lensing},  \cite{diego2020constraining}).}, gravitational microlensing offers the most powerful and robust method of constraining PBHs. In the case of Kepler observations of Galactic sources (\cite{griest2014experimental}), a limit in the planetary mass range: $f(M) < 0.3$ for $2 \times 10^{-9} M_{\odot} < M < 10^{-7} M_{\odot}$ was found, while the MACHO project established that $f(M) \sim 0.1$  in the $[0.1, 1] M_{\odot}$ mass range  (\cite{tisserand2007limits}, \cite{alcock2000macho}).
Mediavilla et al. (\citeyear{mediavilla2017limits}),  based on quasar microlensing observations, have concluded that the fraction of mass in black holes, or any type of compact objects, is negligible outside of the $[0.05,0.45]\, M_{\odot}$ mass range and amounts to $20 \%$ of the total mass in the lens galaxies (\cite{mediavilla2017limits}). Consequently,  the existence of a significant population of intermediate-mass primordial black holes (IMPBH) appears to be almost inconsistent with current microlensing observations.

Recently, Niikura et al.  (\citeyear{niikura2019microlensing}), using the largest sample of microlensing events for stars in the Galactic bulge (2622 microlensing data) obtained from the 5-years OGLE observation, were able to constrain the PBH
abundance  in the Milky Way (MW). They showed that the 6 observed ultrashort-timescale events can be well explained by PBHs with mass about the Earth-mass, provided that these PBHs constitute about $1\%$ of the DM in the MW.  Moreover, from a seven hour lasting  observation with the Subaru Hyper Suprime-Cam (HSC) towards the  $M31$ galaxy, a single candidate event was identified,  which translates into a rather stringent upper bounds on the abundance of PBHs in this mass range    $[10^{-11}, 10^{-6}]~M_{\odot}$ (\cite{niikura2019constraints}. However, we caution that the statistics is very low and one should not rely too much on the above results.

Until now, several microlensing surveys have been undertaken towards the Galactic bulge (in particular $MOA$ and
$OGLE$), $M31$ ($MEGA$ (\cite{ingrosso2007new})), $AGAPE$  (\cite{belokurov2005point}), $WeCAPP$ (\cite{riffeser2008m31}), $PLAN$  (\cite{novati2014m31}), and the $Angstrom$ Project (\cite{kerins2006angstrom}) and the Large and the Small Magellanic Clouds (in particular $MACHO$, $EROS$   (\cite{tisserand2007limits}), $SuperMACHO$   (\cite{rest2005testing}). However, due to the way these surveys were designed, they were not able to give reliable constraints on PBHs in the Galactic halo.

Euclid is  a space-based mission, planned mainly for weak lensing observations of cosmological interest (\cite{laureijs2011euclid}).   An additional survey of the Euclid telescope, although not definitely approved as yet, is to observe  for a few months the Galactic bulge.  The galactic coordinates of the Euclid line of sight toward the Galactic bulge center are $ l = 1.1^\circ,   b = -1.7^\circ $. Euclid is a Medium Class mission of the ESA (European Space Agency), scheduled to be launched in $2022$ and planned to be put at the L2 Lagrangian point  (\cite{racca2016euclid}).


The purpose of this paper is the investigation of the traces of the Galactic halo PBHs (if PBHs contribute to the Galactic halo DM) via microlensing observations. Since PBHs are expected to be  distributed from the Galactic center up to  the outer halo region, due to their large velocity dispersion they can cause microlensing events not only towards the Magellanic Clouds $(LMC, SMC)$ and $M31$, but also towards the Galactic bulge.  We calculate the optical depth and microlensing rate towards these four targets considering a wide range of possible masses for the PBHs.  Also, we forecast the number of events caused by PBHs and observable by the Euclid telescope in the case that the satellite will be used for microlensing searches at the end of its nominal mission.

The structure of this paper is as follows: in Section 2 we briefly review the basics of gravitational microlensing and give the relevant equations  for the optical depth and microlensing rate calculations. In Section 3 we review the standard models adopted for the mass density of the MW halo and for the source stars in the MW bulge, LMC, SMC and M31. In section 4 we give and discuss the main results about the optical depth and microlensing rate, focusing on the Euclid observations. Finally, our conclusions are drawn in Section 5.

\section{Microlensing  by PBHs} 
Since the gravitational microlensing phenomenon does not rely on the flux output from the lensing object, it is one of the few methods of finding primordial black holes (PBHs), if they exist. These dark compact objects, which may spread over a wide range of masses, may populate the Milky Way halo and microlens background stars. For definitness, and in order to investigate the capability to reveal PBHs in different directions, i.e towards the Galactic Bulge, the Magellanic Clouds and the M31 galaxy, through microlensing observations, we assume that a fraction  $f$ of the Galactic halo dark matter is in the form of PBHs.

\subsection{Microlensing basics } 
A microlensing event occurs when a compact object (lens) approaches very closely to the observer’s line of sight toward a background source star (\cite{paczynski1986gravitational}).  In the simplest case, when the point-like approximation for both the lens and the source holds, \footnote{In our analysis we also assume that the relative motion among the observer, lens, and source is uniform (with constant velocity) and linear (on a straight line). See the discussion on this issue in References  (\cite{paczynski1996gravitational},  \cite{gould2000natural})  and references therein.} individual images cannot be resolved due to their small separation, but the total brightness of the images increases  with respect to that of the unlensed source, leading to a specific   time-dependent amplification of the source star brightness, that is
\begin{equation}
    A=\frac{2+u^2}{u \sqrt{4+u^2}}.
    \label{A}
\end{equation}
Here, $u(t)$  is the distance, in units of the Einstein radius, between the source star and the lens at an observation epoch $t$, and is given by
\begin{equation}
    u(t)=\sqrt{u^2_0+ \left(\frac{t-t_0}{t_E}\right)^2},
    \label{u(t)}
\end{equation}
where $t_0$ and $u_0$ are the time and impact parameter at the closest approach, whereas $t_E=R_E/v_T$ is the Einstein timescale, which is defined as the time required for the lens to transit the Einstein radius.
The Einstein radius is the radius of the ring image formed when the observer, the lens and the source are perfectly aligned. It is given by, $R_E=\sqrt{\frac{4GMD_S}{c^2}x(1-x)}$, where $M$ is the mass of the lens and $x=D_L/D_S$  is the normalized lens distance, whereas $D_S$  and $D_L$   are the source-observer and lens-observer distances, respectively.  We also note that $v_T$  is the relative transverse velocity between the lens and the source.
During a microlensing event, the source position projected in the lens plane encounters the Einstein ring, where the projected separation is $u_T = 1$. From eq.  (\ref{A}) one easily gets that this condition corresponds to a threshold value for the source amplification which  takes the value $A_T = 1.34$. We note that, in particular in the case of space-based observations, due to the absence of seeing effects, the amplification threshold may be even much smaller than $1.34$, with a corresponding much larger value of $u_T$.  In the case of the Euclid telescope, for example, the photometric error  is expected to be less then  $\sim 0.1 \% $ (see, e.g., \cite{regnault2013photometric}), which leads to a threshold amplification $A_T = 1.001$. From eq. (\ref{A}) one get the corresponding value of $u_T \simeq 6.54$, which can be considered as an upper limit for the impact parameter (\cite{griest2011microlensing}, \cite{hamolli2013theoretical}). 
By photometric measurements of the event lightcurve, three parameters can be defined: $t_0$, $t_E$ and $u_0$. However, among these parameters, only $t_E$ contains information about the lens and this gives rise to the so-called parameter degeneracy problem, which does not allow to infer the lens parameters uniquely. To break this degeneracy,  second-order effects, as the finite source effects and the parallax effect, can be considered. Also, it is well known that a gravitational microlensing event gives rise to an astrometric shift of the source, which may be extremely useful to break, at least partially, the parameter degeneracy problem in microlensing observations (\cite{hamolli2019free}, \cite{nucita2017astrometric}).

\subsection{The optical depth}

The microlensing optical depth $\tau$  is the probability that any given star is being significantly lensed at any given time by a foreground compact object. For a  star and a given model of the lens density distribution one can calculate  $\tau$ through the equation
\begin{equation}
    \tau=\int_0^{D_S}n(D_l)\pi u_T^2R_E^2dD_l
    \label{tau}
\end{equation}
which is just the probability that a lens object is inside the microlensing tube (\cite{paczynski1986gravitational},  \cite{gaudi2012microlensing}). Here, $n=\rho/M$ is the number density of the lenses  ($\rho$   is their mass density) and  $M$ is the lens mass. Since  $R_E^2\propto M$, it turns out that the optical depth is simply a geometrical quantity and does not depend on the lens mass. By using the definition of the Einstein ring radius, and scaling the distance along the line of sight to the distance to the source, one arrives at a relatively simple expression for the optical depth:
\begin{equation}
    \tau(D_S)=\frac{4\pi u_T^2GD_S^2}{c^2}\int_0^1\rho(x)x(1-x)dx.
    \label{tauDS}
\end{equation}
An important note at this point is that the sources are not all at the same distance $D_S$, but are distributed along the line of sight. Therefore, a more correct derivation of the optical depth expression should take into account the distribution of the source stars. We then should integrate not only over the distance of the lenses, but also on the distance of the source stars (see  \cite{kiraga1994gravitational}) and  define the average optical depth as:

\begin{equation}
<\tau>=\left[ \int_{D_{S,min}}^{D_{S,max}}dD_S n(D_S) D_S^2\right]^{-1}\int_{D_{S,min}}^{D_{S,max}}dD_S n(D_S) D_S^2 \tau(D_S)
\label{<tau>}
\end{equation}

where $D_{S,min}$ and $D_{S,max}$ are the minimum and maximum  distances (up to the boundary of the observation region) from the observer's position and $n(D_S)$  is the number density distribution of source stars.
If we consider the above mentioned  scenario, that is that PBHs constitutes some mass fraction $f$ of the halo $DM$ in the $MW$ halo region, where $f=\frac{\Omega_{PHB}}{\Omega_{DM}}=\frac{\rho_{PHB}}{\rho_{DM}}$, the  optical depth due to the PBHs can be written in the form $<\tau>_{PBH}=f<\tau>_{DM}$  .

\subsection{The microlensing rate}

Somewhat more interesting and important in microlensing observations is the event rate, which is the rate at which a given background star undergoes a microlensing event due to a foreground lensing object. The microlensing rate is calibrated to show the probability per year that a source star is microlensed.
The microlensing event rate for any given source is extremely small, but it can be maximized toward regions with a high surface density of sources, as the Galactic bulge, $LMC$, $SMC$ and the $M31$ galaxy. Of course, the lensing objects are constituted by the populations of compact objects along the line of sight towards those regions.

In order to define the microlensing rate, we consider the geometry and variables defined in Fig. 4 of  Griest (\citeyear{griest1991galactic}), which gives the differential event rate ($d \Gamma$) of a lensing object entering a volume element along the line-of-sight, where the lens causes a microlensing with magnification above a certain threshold value.  Integrating  it along the line of sight to the source,  the microlensing rate per background star is defined as
\begin{equation}
\Gamma = \int d \Gamma= \int \frac{n(D_l)f(\vec{v})d^3x d^3v}{dt},
 \label{gamma}
\end{equation}
where $n(D_l)$  is the number density of the lenses and $f(\vec{v})$ is the lens velocity distribution. The volume element, $d^3x$,  can be written  as $d^3x=v_{\perp}cos(\theta)dtR_Eu_Td\alpha d(D_l)$,  where $v_{\perp}$ is transverse velocity, $R_E$ is the Einstein radius, while $\theta$ is the angle between $v_{\perp}$ and the
normal to the lateral superficial element, $dS = dldD_l$, of
the microlensing tube, with $dl = R_E~du_T~d\alpha$ being the cylindrical segment of the tube  (\cite{jetzer2002microlensing}). Since the component of the lens velocity parallel to the line of sight does not enter into the description of the microlensing phenomenon, it is convenient to write $\vec{v}=\vec{v_{\perp}}+\vec{v_{\parallel}}$  and the velocity element as $d^3v=d^2v_{\perp}dv_{\parallel}$. The perpendicular component of the velocity element in cylindrical coordinates can be then written as $d^2v_{\perp}=v_{\perp} dv_{\perp} d\theta$. Therefore, eq. (\ref{gamma}) can be rewritten as:
\begin{equation}
    d\Gamma=\frac{\rho}{M} u_T R_E dD_lf(v_{\perp},v_{\parallel}) d^2v_{\perp}dv_{\perp} d\theta \cos(\theta) d\alpha dv_{\parallel}.
    \label{gamma1}
\end{equation}
We note that the parameters $\theta, \alpha, v_{\perp}$ in the equation above vary in the range  $ [-\pi/2, \pi/2]$, $ [0, 2\pi]$  and $ [0, \infty)$, respectively. Considering the velocity distribution of the lenses to be a  Maxwellian distribution with dispersion velocity $\sigma_l$ one gets
\begin{equation}
    f(\vec{v})d^3v=\frac{1}{\pi^{3/2} \sigma_l^3} exp(-\frac{v_l^2}{\sigma_l^2})d^3v.
    \label{dist}
\end{equation}
The velocity distribution given above can be further simplified as $f(v_{\perp},v_{\parallel})=f(v_{\perp})f(v_{\parallel})  $. It also satisfies the normalization condition  $\int d^2v_{\perp} \int dv_{\parallel} f(v_{\perp},v_{\parallel})=1$.  So, by integrating eq. (\ref{dist}) on the parallel velocity component, one can obtain the transverse velocity distribution (\cite{jetzer2002microlensing}):
\begin{equation}
    f(v_{\perp})=\frac{1}{\pi \sigma_l^2} exp(-\frac{v_{l\perp}^2}{\sigma_l^2}).
    \label{distribution1}
\end{equation}
The transverse component of the lens velocity can be written as $\overrightarrow{v_{l\perp}}=\overrightarrow{v_{\perp}}+\overrightarrow{v_{t\perp}}$, where $v_{\perp}$ is the lens transverse velocity related to the line of sight and  $v_{t\perp}$  is the transverse velocity of the microlensing tube. We note that $v_{t\perp}$  is related to both  $v_{s\perp}$ and  $v_{\odot\perp}$, which  are the transverse velocities of the source star  and of the observer, respectively \footnote{We note that the observer is assumed to be co-moving with the Sun.}.  Then, one gets the relation
$\overrightarrow{v_{t\perp}}=(1-D_L/D_S)\overrightarrow{v_{\odot\perp}}+(D_L/D_S) \overrightarrow{v_{S\perp}}$. As usual, we calculate the transverse tube velocity by $v_{t\perp}^2=(1-x)v_{\odot\perp}^2+x^2v_{S\perp}^2 +2x(1-x)v_{\odot\perp}v_{S\perp} cos \beta $, where
$\beta$  is the angle between $\overrightarrow{v_{S\perp}}$ and $\overrightarrow{v_{\odot\perp}}$. As far as the source stars are concerned,  we also assume that their velocity distribution is Maxwellian with velocity dispersion, $\sigma_s$. Therefore we have
\begin{equation}
  f(v_{s\perp})d^2v_{s\perp}=\frac{v_{s\perp}}{\pi \sigma_s^2} exp(-\frac{v_{s\perp}^2}{\sigma_s^2})dv_{s\perp}d\phi
    \label{distribution}
\end{equation}
where $\phi$ is is the angle between $\overrightarrow{v_{s\perp}}$ and $\overrightarrow{v_{\perp}}$.

Taking all that into account, the microlensing rate is given by

\begin{equation}
\begin{split}
\Gamma(D_S) & = \frac{2u_TD_S}{M}\int_0^1 dx \rho(x) R_E(x) \times \int_0^{\infty} \frac{v_{s\perp}}{\pi \sigma_s^2} exp(-\frac{v_{s\perp}^2}{\sigma_s^2})dv_{s\perp} \\
& \times  \int_0^{\infty} \frac{v^2_{\perp}}{\pi \sigma_l^2} exp(-\frac{v_{\perp}^2+v_{t\perp}^2}{\sigma_l^2})dv_{l\perp} \times \int_0^{2\pi}  exp(-\frac{2v_{\perp}v_{t\perp}cos\beta}{\sigma_l^2}) d\beta \times \int_0^{2\pi} d\phi.
\end{split}
\label{gamma2}
\end{equation}

Using variables: $ y=v_{l\perp}/\sigma_l$, $ \eta=v_{t\perp}/\sigma_l$ and $ z=v_{s\perp}/\sigma_s$, the microlensing rate is

\begin{equation}
\begin{split}
\Gamma(D_S) & =4u_T D_S\sqrt{ \frac{4GMD_s}{c^2}}\frac{\sigma_l}{\pi M}\int_0^1 dx \rho(x) \sqrt{ x (1-x)} \times \int_0^{\infty} z exp(-z^2)dz \\
& \times  \int_0^{\infty}y^2 exp(-y^2-\eta^2)dy
\times  \int_0^{2\pi}  exp(-2y\eta cos\beta) d\beta.
\label{gammaDS}
\end{split}
\end{equation}
Finally, the average microlensing rate  per star can be rewritten as (see, e.g.,  \citealp{kiraga1994gravitational} for further details)
\begin{equation}
<\Gamma>=\left [\int_{D_{S,min}}^{D_{S,min}}dD_S n(D_S) D_S^2\right]^{-1}\int_{D_{S,min}}^{D_{S,min}}dD_S n(D_S) D_S^2 \Gamma(D_S),
\label{<gama>}
\end{equation}

In the same way, the  microlensing rate  due to the PBHs can be written in the form $<\Gamma>_{PBH}=f<\Gamma>_{DM}$. 

\section{Stellar and dark matter distributions in the Milky Way}

We trace the microlensing events caused by PBHs in Galactic halo towards some specific directions, which appear to be the most favorable directions for actual observations: the $MW$ bulge, $LMC$, $SMC$ and the $M31$ galaxy. Below we provide some details for the mass density of the halo dark matter and the model distribution of the stellar components.

\subsection{PBH lenses}

We consider a scenario in which PBHs constitute some mass-fraction $f$ of the DM component in the MW halo region and  assume that the PBH distribution follows the standard halo model (S), which consists of a cored isothermal sphere with dark matter density given by:
\begin{equation}
    \rho_{DM}(r)=\rho_0\frac{r_c^2 +r_0^2}{r_c^2 +r^2},
    \label{lens}
\end{equation}
where $r$ is the galactocentric radius, $\rho_0\simeq 0.0079\, M_{\odot}pc^{-3} $  is the local dark matter density, $r_c \simeq 5.6$ kpc is the halo dark matter core radius and $ r_0 \simeq 8.5$ kpc  is the  galactocentric distance of the Sun  (\cite{green2017astrophysical}). Therefore, the PBH mass density distribution is  $\rho_{PHB}(r)=f~\rho_{DM}(r)$.

\subsection{Source star distribution in the Galactic bulge}

For the mass density distribution of the stellar population in the Galactic bulge  we use the triaxial bulge model given by
\begin{equation}
    \rho_b=\frac{M_b}{8\pi abc} exp\left(-\frac{s^2}{2}\right).
    \label{lens1}
\end{equation}
Here $s^4= (x^2/a^2+y^2/b^2)^2+z^4/c^4$, the scale length values  are $a=1.49$ kpc, $b=0.58$ kpc and $c=0.40$ kpc (we remind that  the coordinates $x$ and $y$ span the galactic disk plane, whereas $z$ is perpendicular to it) and the bulge mass is $M_b\approx 2\times 10^{10}\,  M_{\odot}$ (\cite{hafizi2004microlensing}).

\subsection{Source star distribution in the LMC}

The source star distribution in the LMC is generally described  as a luminous disk and a bar  (\cite{jetzer2002microlensing}). The distance between the center of the LMC and the observer is $D_{LMC}\simeq 50$ kpc and the stellar disk is modeled with a double exponential profile:
\begin{equation}
    \rho_{disk}=\frac{M_d}{4\pi z_d R^2_d} \exp(-R/R_d-|z|/z_d),
    \label{LMCdisk}
\end{equation}
where $R=\sqrt{x^2+y^2}$. $R_d\simeq 1.6$ kpc  is the radial scale length, $z_d\simeq 0.3$ kpc is the vertical scale  height  and $M_d=[2.4, 4.8]\times 10^9\, M_{\odot}$ is the LMC disk mass. The galactic coordinates for the center of the LMC disk are   $l = 297.7^\circ ,\, b = -33.5^\circ$. The LMC disk is inclined by an  angle $ i=30^\circ $  with respect to our line of sight and its position angle is $\phi= 170^\circ$.  In addition to the disk, the LMC possess a stellar bar described by  a triaxial Gaussian density profile:
\begin{equation}
    \rho_{bar}=\frac{M_{bar}}{(2\pi)^{3/2} x_by_bz_b} \exp\left(-\frac{1}{2}\left[\left(\frac{x}{x_b}\right)^2+\left(\frac{y}{y_b}\right)^2 +\left(\frac{z}{z_b}\right)^2\right]\right),
    \label{LMCbar}
\end{equation}
where $x$, $y$, $z$ are coordinates along the principal axes of the bar. Here, $x_b=1.0 $ kpc, $y_b=0.3$ kpc, $z_b=0.3$ kpc are the scale lengths along the three axes and $M_{bar}= [0.6, 1.2] \times 10^9M_{\odot} $ is the total mass of the bar. The bar is inclined by an angle $ i=30^\circ $  and has a position angle  $\phi= 120^\circ$.  The galactic coordinates for the center of the  LMC stellar bar are    $l = 280.5^\circ , b = -32.8^\circ$.

\subsection{Source star distribution in the SMC}
The SMC is a dwarf irregular galaxy orbiting the MW and in tight interaction with the LMC. Accordingly to the star formation history of the SMC, two stellar components can be distinguished: an old star (OS) and a young star (YS) population (see \cite{calchi2013microlensing}). For the YS mass distribution, we adopt a spheroidal model with a fully Gaussian profile:
\begin{equation}
    \rho_{SMC}^{(YS)}=\rho_0^{(YS)} \exp\left(-\frac{1}{2}\left[\left(\frac{\xi}{\xi_{YS}}\right)^2+\left(\frac{\eta}{\eta_{YS}}\right)^2 +\left(\frac{\zeta}{\zeta_{YS}}\right)^2\right]\right).
    \label{SMC(YS)}
\end{equation}
As far as the OS mass distribution is concerned, we keep the Gaussian profile along the line of sight, and a smoother exponential profile in the orthogonal plane:
\begin{equation}
    \rho_{SMC}^{(OS)}=\rho_0^{(OS)}
    \exp\left[-\sqrt{ \left(\frac{\Xi}{\Xi_{OS}}\right)^2+\left(\frac{\Upsilon}{\Upsilon_{OS}}\right)^2}\right]\exp\left[-\frac{1}{2}\left(\frac{Z}{Z_{OS}}\right)^2\right].
    \label{SMC(OS)}
\end{equation}
The central density values for the YS and OS population are $\rho_0^{(YS)}=8.5 \times 10^6\, M_{\odot}\, kpc^{-3}$  and $\rho_0^{(OS)}=3.9 \times 10^7\, M_{\odot}\, kpc^{-3}$, respectively.  The value of the position angle, with respect to the north direction, is fixed at $\phi_{YS}= 66^\circ$ and $\phi_{OS}= 83^\circ$  for the YS and OS populations, respectively. The reference frames $(\xi,\eta, \zeta )$ and $(\Xi,\Upsilon, Z )$  are directed along the principal axes of the YS and OS spheroid, respectively. For the YS population, we fix $(\xi_{YS},\eta_{YS},\zeta_{YS} )=(0.8, 3.5, 1.8)$ kpc  and for the OS population, $Z_{OS} = 2.1$ kpc,  $(\Xi_{OS},\Upsilon_{OS} )=(0.8, 2.1)$ kpc. The galactic coordinates for the SMC are $l = 307^\circ , b = -46^\circ$.  The YS population is inclined by the angle $ i=74^\circ $, while the OS population  is not  inclined.  The OS and YS populations have a relative distance shift between them, since the centre of the YS population is at a distance of about $2$ kpc behind that of the OS component, which is at a distance of   $\simeq 63.5$ kpc from the Earth.

\subsection{Source star distribution in the M31 galaxy}
The morphology of M31 is similar to that of the Milky Way, with a central bulge and a disc. The distance between the center of the M31 and the observer is $D_{M31}\simeq 778$ kpc.  The M31 bulge is parameterized by a flattened power law of the form
\begin{equation}
    \rho_B(R,z)=\rho_B(0)\left[1+(\frac{R}{a})^2+ q^{-2} (\frac{z}{a})^2)\right]^{-s/2},
    \label{M31(B)1}
\end{equation}
where the coordinates $x$ and $y$ span the M31 disk plane ($z$ is perpendicular to it),  $\rho_B(0)\simeq 4.5\times 10^9 M_{\odot} kpc^{-3} $, $q\simeq 0.6$ is the ratio of the minor to major axes, $a\simeq 1.0$ kpc  is the core radius and  $s=3.8$   is the power-law index (\cite{de2005influence}). The mass density of the M31 disk stars adopting the parameters of the Reference model in (\cite{kerins2004impact}) is described by a sech-squared profile as follows:
\begin{equation}
    \rho_D(R,z)= \rho_D(0)\exp (-R/h) \sech^2 (z/H),
    \label{M31(B)2}
\end{equation}
where $R$ is the distance to the center of the M31 disk, $H\simeq 0.3$ kpc and $h\simeq 6.4$ kpc  are, respectively, the scale height and scale lengths of the disk and $\rho_D(0)\simeq 3.5\times 10^8\, M_{\odot} kpc^{-3} $  is the central mass density. The M31 disk is inclined by the angle  $ i=77^\circ$  to our line of sight and has position angle $\phi= 38.6^\circ$.

\section{Model results}
Let us now consider that a fraction $f$ of the Galactic halo DM is in the form of PBHs, assumed to have a mass density distribution exactly as that of the halo DM.

First, we consider a monochromatic mass distribution for the PBHs, so we assume that all PBHs have the same mass. 
We calculate the optical depth and the microlensing rate for this model, varying masses in the range $10^{-14}\, M_{\odot} \leq  M \leq 10^2\, M_{\odot}$ (see, e.g. Sureda et al. 
(\citeyear{sureda2020press})  and references therein). 
Using the Galactic celestial coordinate variables $(l, b)$, an object at the distance $d$ from the observer (Sun's position) is at a distance $r=\sqrt{r_0^2-2r_0 d \cos l \cos b +d^2}$ from the Galactic center, where $r_0\simeq  8.5$ kpc is the distance between the Sun and Galactic center, $d=\sqrt{x'^2 + y'^2 +z'^2}$  and $x'=d \cos b \cos l$,  $y'=d \cos b \sin l $,  $z'=d \sin b$.  
In the Cartesian  coordinate system $(x', y', z')$, the observer is at the coordinate origin.
Since we are interested in estimating the optical depth and microlensing rate due to the PBHs in the Galactic halo towards three nearby galaxies, i.e.  $LMC$, $SMC$ and $M31$, we introduce the coordinate system $(x_0, y_0, z_0)$ , which has the origin in the center of these galaxies at $(l_0, b_0, D_0)$ and has the $z_0-axis$ directed  toward the observer: the $x_0-axis$ is assumed to be anti parallel to the galactic longitude axis and the $y_0-axis$ parallel to the galactic latitude axis. For an object at distance $D$ from the observer, a point with galactic coordinates $(l, b)$ can be defined through the  coordinates $(x_0, y_0, z_0)$ as:
\begin{equation}
    \label{eq:t}
    \begin{split}
        & x_0=- D \cos b \sin(l-l_0)\\
        & y_0= D \sin b \cos b_0 -D \cos b \sin b_0 \cos(l-l_0)\\
        & z_0= D_0- D \cos b \cos b_0 \cos(l-l_0) -D \sin b \sin b_0 \\
    \end{split}
\end{equation}
Therefore, in  order to find the coordinates $(x,y,z)$ according to the selected systems, in the case of LMC and M31 they are obtained by rotating around $z_0-axis$ with position angle $\phi$ counterclockwise and around the new $x-axis$ with the inclination angle $i$  clockwise. In the case of SMC, the coordinates  $(\Xi,\Upsilon, Z )$ for the  OS population and $(\xi,\eta, \zeta )$ for the YS population  are obtained by rotating around $z_0-axis$ with the position angle  $\phi$ counterclockwise and around the new $x-axis$ with the inclination angle  $i$ counterclockwise.

\subsection{Optical depth}
\begin{figure}[t]
\plotone{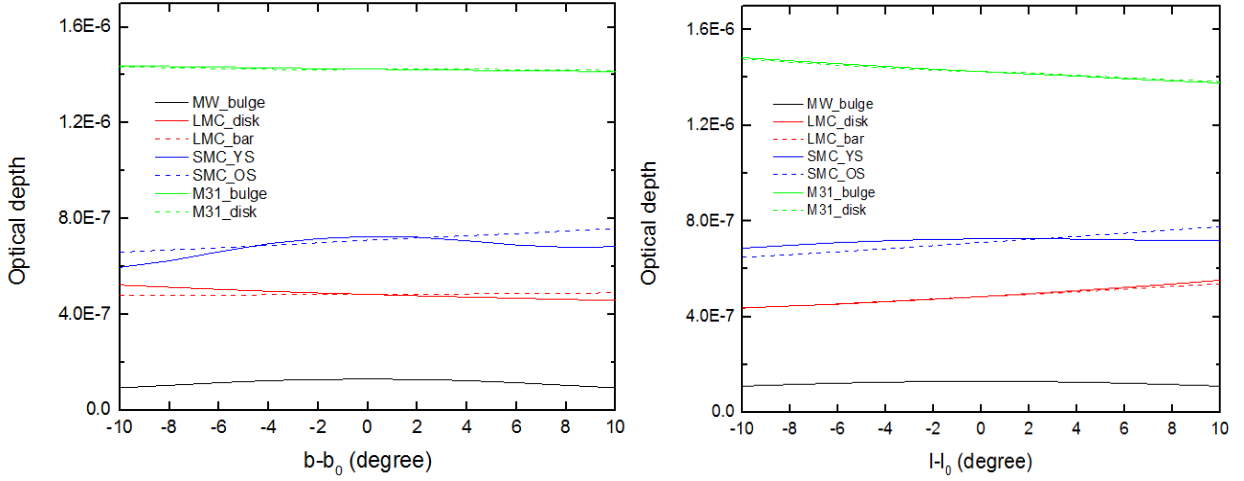}
\caption{Left panel: optical depth curves towards the MW bulge (black line), LMC (red line), SMC (blue line), and M31 (green line) as a function of the latitude $b-b_0$. The value of the longitude is set to be $l_0$. Right panel: the optical depth for the same target objects as in the left panel, depending on the longitude $l-l_0$. The value of the latitude is set to $b_0$. The considered values of  $(l_0, b_0)$  are:  for the Galactic bulge  $(l_0, b_0) = (0^\circ, 0^\circ)$,  for LMC  $(l_0, b_0) = (280.5^\circ, -32.8^\circ)$, for SMC $(l_0, b_0) = (307^\circ, -46^\circ)$, and  for M31  $(l_0, b_0) = (121.17^\circ, -21.57^\circ)$.  }
\label{fig1}
\end{figure}

Based on eq. (\ref{<tau>}) we estimate the average optical depth due to the PBHs towards the Galactic bulge, LMC, SMC and the M31 galaxy. In Fig.\ref{fig1} the optical depth due to PBHs is given towards the Galactic bulge, LMC, SMC and M31 galaxy as a function of the latitude $b - b_0$ (with the longitude fixed at the value $l_0$), and as a function of the longitude $l-l_0$ (with the latitude fixed at $b_0$). The $(l_0, b_0)$ values of the center for considered galaxies are: $(l_0, b_0) = (0^\circ, 0^\circ)$ for the Galactic bulge, $(l_0, b_0) = (280.5^\circ, -32.8^\circ)$ for LMC,  $(l_0, b_0) = (307^\circ, -46^\circ)$ for SMC and  $(l_0, b_0) = (121.17^\circ, -21.57^\circ)$ for M31.
For definiteness, the plots are given for $f_{PBH} = 1$ and $u_T=1$. In case of other values, the rescaling is obviously a multiplication by $f_{PBH}$ and by $u_T^2$.

As one can see from Fig.\ref{fig1}, the largest values of the optical depth are found  towards M31 $(\tau \sim 10^{-6})$, with a factor about 10 times larger than those towards the Galactic bulge. This can be certainly attributed to  the enormous volume and large mass content between the Earth and M31.
 Indeed, microlensing observations towards M31 are able to allow a full map of the Galactic halo, whereas the other considered targets are immersed within the Galactic halo.
One can also remark that the optical depth towards SMC is significantly larger than that towards LMC. This has to be attributed mainly to the larger value of the SMC center longitude  with respect to the corresponding value of LMC, but also to its larger distance. The LMC optical depth is about $5 \times 10^{-7}$, in agreement to the result in \cite{griest1991galactic}, in which the adopted values of the model parameters are slightly different.  We also remark that the optical depth towards the Galactic bulge is slightly dependent on the direction and keeps a value about $10^{-7}$, with a slender maximum towards $l_0,b_0$. This is compatible with the results obtained in \cite{niikura2019microlensing} (see e.g Table I).  In the case of LMC, the optical depth decreases with the latitude and increases with the longitude. Concerning SMC, there is a more visible difference between the role of the two star populations, mainly  due to the fact that their centers are shifted by about $2$ kpc one with respect to the other.  The ratio of the optical depth towards  $SMC$ and $LMC$  turns out to be  $<\tau>_{SMC} ~\simeq ~1.4  ~<\tau>_{LMC}$, consistent with the results to \cite{afonso2003limits}. The optical depth towards the M31 galaxy is almost constant with respect to the latitude and decreases appreciably by increasing the longitude.  We also notice that the obtained values are in agreement  with the results obtained in (\cite{niikura2019constraints}).

As far as the planned  Euclid's observations towards the Galactic bulge is concerned,  the microlensing optical depth by PBHs turns out to be  $<\tau>\simeq 1.30 \times 10^{-7}$.

\subsection{ Microlensing rate }

Based on eq. (\ref{<gama>}), we estimate  the average microlensing rate, considering the same conditions as in the previous section. For definiteness and to simplify the discussion  we assume that all the PBHs have the same mass, $M=1 \,M_{\odot}$, but the reader should keep in mind that the rate scales as $\Gamma\propto M^{-1/2}$ so that it is straightforward to rescale the results for a different PBH mass value. Concerning the transverse velocity dispersion, we keep the value $\sigma_l=210$ km/s as it is considered for the lenses in the MW halo (\cite{jetzer2002microlensing}). For the source star populations in the MW bulge, the velocity dispersion is assumed to be $\sigma=156$ km/s, whereas in the case of the  M31 galaxy, we consider $\sigma_{bulge}=100$ km/s and $\sigma_{disk}=30$ km/s (\cite{de2005influence}). For  LMC, we have $\sigma_{disk}=20.2$ km/s and $\sigma_{bar}=24.7$ km/s,  while for SMC, $\sigma_{YS}=20$ km/s and  $\sigma_{OS}=30$ km/s (\cite{novati2006microlensing}).

\begin{figure}[t]
\plotone{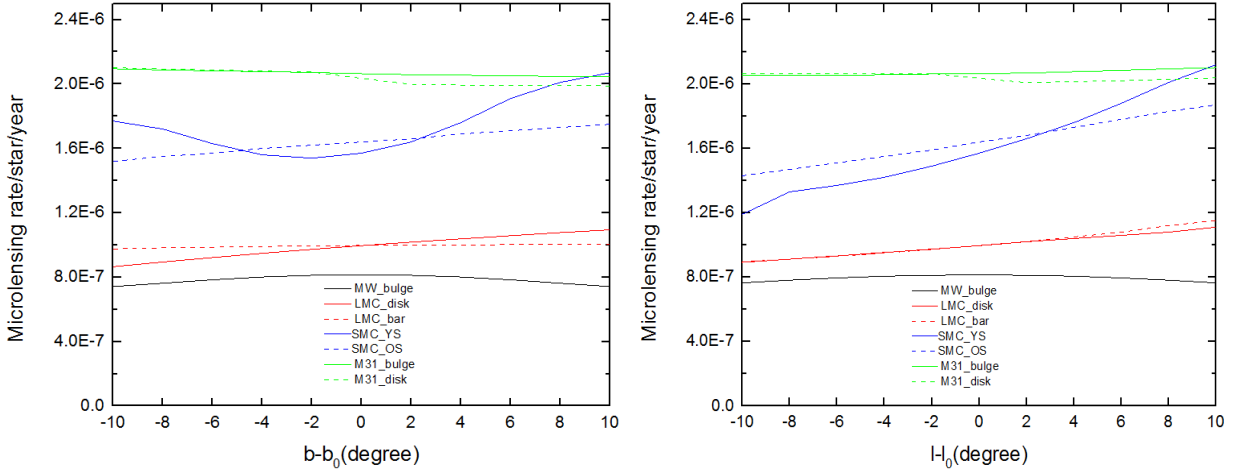}
\caption{Left panel: The microlensing rate caused by our Galactic halo PBHs to the stars in MW bulge (black line), LMC (red line), SMC (blue line), and M31 (green line) is plotted  as a function of the latitude $b-b_0$ (the longitude value is set here to $l_0$). Right panel: the microlensing rate is plotted  as a function of the longitude $l-l_0$ (the latitude is fixed here to $b_0$). }
\label{fig2}
\end{figure}

In Fig. \ref{fig2} we plot the PBH  microlensing rate per star per year towards the MW bulge, LMC, SMC and M31 as a function of the latitude (with the longitude  fixed at $l_0$), and as a function of the longitude (with the latitude fixed at $b_0$). The values of  $(l_0, b_0)$ are the same as before.  We also remind that the plots are given for $f_{PBH} = 1$ and $u_T=1$. In case of other values, the rescaling is obviously obtained with a multiplication by $f_{PBH}$ and  $u_T$.

As one can see, the  highest value of the microlensing rate is obtained in the case of observations towards the M31 galaxy (except for some specific directions for the stars in SMC, with farthest longitudes and latitudes) and this has to be attributed to the dark matter distribution function as a cored isothermal sphere. However, the observable rate of events may be smaller because it depends on the number of observed stars that, in particular towards M31, is smaller due to its larger distance. 

As far as the microlening rate towards the Galactic bulge is concerned, it peaks, obviously, at the bulge center and has a value of about $8\times 10^{-7}$. In the case of the LMC, the same figure shows a slightly increasing trend both as a function of the latitude and longitude values. In the case of the $SMC$, the microlensing rate has a somewhat irregular profile due to the influence of the two different stellar populations. Concerning the stars in the  $M31$ galaxy, the microlensing rate has a almost constant behaviour with average value about $2\times 10^{-6}$.

In Table \ref{table1} we present  the  average microlensing rate, obtained from Eq. \ref{<gama>},  for events  caused by PBHs towards the center of the MW bulge, LMC, SMC and M31, keeping the same value for the PBH mass, $1\, M_{\odot}$.  The scaling  in relation to the $f_{PBH}$ and $u_T$ is simply multiplying by these factors.
\begin{table}[htbp!]
    \centering
    \begin{tabular}{|c|c|}
        \hline
        $ $           & $<\Gamma> (M=1\, M_{\odot})$ \\
                      & $(\times 10^{-7})$           \\
        \hline
        $MW_{bulge}$  & $8.14$                       \\
        \hline
        $LMC_{disk}$  & $9.96$                       \\
        \hline
        $LMC_{bar}$   & $9.97$                       \\
        \hline
        $SMC_{YS}$    & $15.70$                      \\
        \hline
        $SMC_{OS}$    & $16.40$                      \\
        \hline
        $M31_{bulge}$ & $20.60$                      \\
        \hline
        $M31_{disk}$  & $20.40$                      \\
        \hline
    \end{tabular}
    \caption{The average microlensing rate (per source and per year) due to  PBHs with mass $1\,M_{\odot}$. From top to bottom, the rate is given towards the MW bulge, LMC disk and bar, SMC young and old stellar populations and M31 bulge and disk.}
    \label{table1}
\end{table}

The median microlensing timescale is related to the optical depth and microlensing events  \cite{gaudi2012microlensing} through the relation
\begin{equation}
    t_E=\frac{2}{\pi}\frac{\tau}{\Gamma}.
    \label{M31(B)3}
\end{equation}

Since  $\Gamma$ scales with  the PBH mass as $M^{-1/2}$,  the  microlensing timescale $t_E$ turns out to be proportional to the  square root of the PBHs mass. In Table \ref{table2} we give the obtained values for  $t_E$ towards the center of the MW bulge, towards LMC, SMC and the M31 galaxy for the two extremal cases  of $M=10^{-14}\,M_{\odot}$ and $M=10^{2}\,M_{\odot}$ of the PBH mass. As expected, due to the wide mass interval in consideration, the $t_E$ values vary by several orders of magnitude, from a fraction of a second up to several years. For that reason, since a sufficiently high image sampling is requested in order to recognize a microlensing event (let's say at least 5-10 images within $t_E$ are necessary  for that aim), different observing strategies and facilities are needed in order to probe the full PBH mass range.
\begin{table}[htbp!]
    \centering
    \begin{tabular}{|c|c|c|}
        \hline
                      & $t_E (M=10^{-14}\,M_{\odot})$ & $t_E (M=10^2\,M_{\odot})$ \\
                      & (s)                           & (year)                    \\
        \hline
        $MW_{bulge}$  & $0.32$    & $1.02$                    \\
        \hline
        $LMC_{disk}$  & $0.98$    & $3.10$                    \\
        \hline
        $LMC_{bar}$   & $0.98$    & $3.09$                    \\
        \hline
        $SMC_{YS}$    & $0.93$    & $2.95$                    \\
        \hline
        $SMC_{OS}$    & $0.87$    & $2.76$                    \\
        \hline
        $M31_{bulge}$ & $1.39$    & $4.40$                    \\
        \hline
        $M31_{disk}$  & $1.41$    & $4.46$                    \\
        \hline
    \end{tabular}
    \caption{The median microlensing timescale $t_E$ due to halo  PBHs is given for two selected values of the PBH mass $M$, i.e. $M=10^{-14}\,M_{\odot}$ and  $M=10^{2}\,M_{\odot}$. }
    \label{table2}
\end{table}

\subsection{PBHs by Euclid}
Here, we devote a special attention to Euclid survey towards the Galactic bulge (\cite{racca2016euclid}). Since the galactic coordinates of the line of sight are $(l, b) = (1.1^\circ, -1.7^\circ)$ we find the optical depth to be $<\tau>\simeq 1.30 \times 10^{-7}$ and the microlensing rate  per stars per year $<\Gamma>=8.12 \times 10^{-7}$ for the PBH mass, $M=1\, M_{\odot}$. Based on the  magnitude limit of the Euclid telescope, which is foreseen to be  $m=24$,  the mass-luminosity relation  $\frac{L}{L_{\odot}}=(\frac{M}{M_{\odot}})^{3.5}$  and considering the middle point $ 8.5$~kpc away from us, we find the lowest value $M_{min}=0.31M_{\odot}$ of the star's mass to be detected by this telescope. Using a Salpeter mass function  $\frac{dN}{dM}\propto M^{-2.4}$, for stars in the Galactic bulge, we calculate their mean mass value $<M>=0.97\,M_{\odot}$.

Considering the distance of the bulge stars to be within  the range $[7,  10]$ kpc with  mass density distribution given by equation (\ref{lens1}), the Euclid’s field of view of $0.54$ square degree    and the galactic coordinates of the line of sight $(l, b) = (1.1^\circ, -1.7^\circ)$, we obtain that the number $N_{ED}$ of detectable source stars in the Euclid microlensing observations is  $N_{ED} = 6.45\times 10^7$. By multiplying $<\Gamma>N_{ED}t_{obs}$, where $t_{obs}$ is the observation timescale in years, we find  about $340$ microlensing events per year ( for $f_{PBH} = 1$ and $u_T=6.54$),  with a microlensing  timescale  $t_E\sim 8$ months in the case of a PBH mass of $1\, M_{\odot}$.

\begin{figure}[t]
\plotone{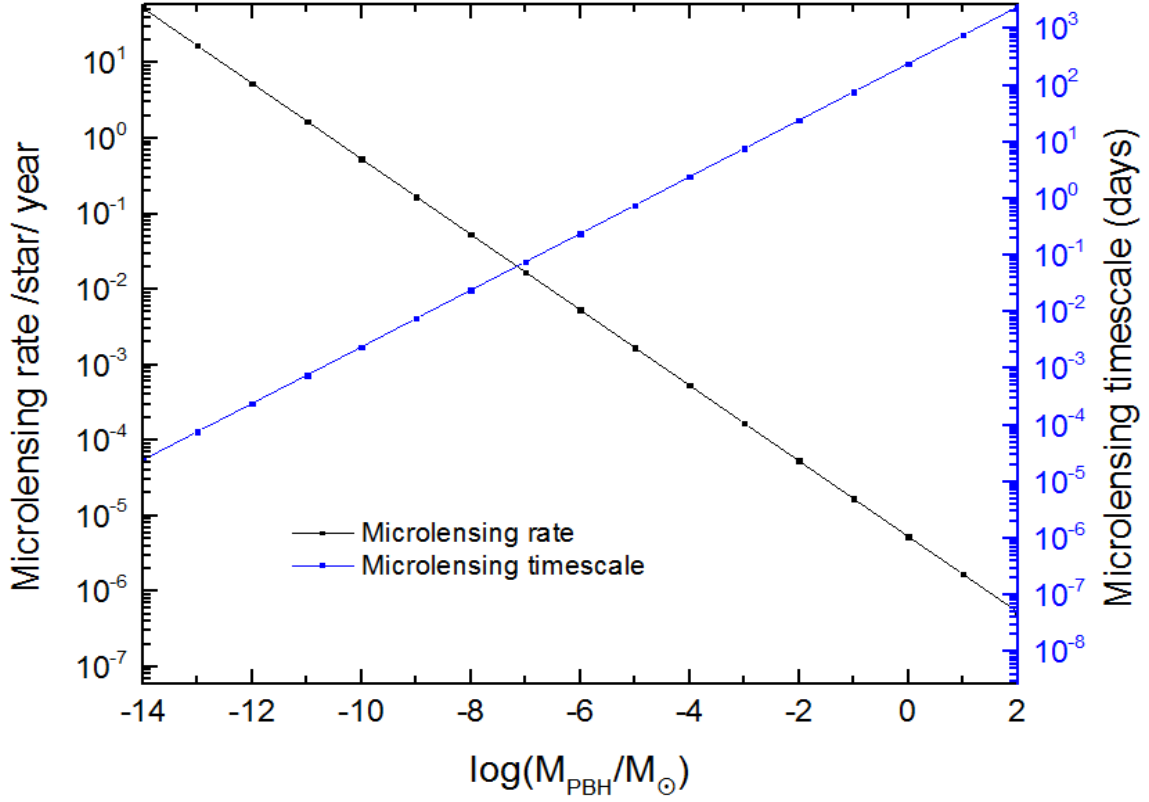}
\caption{The microlensing rate (black line) and  the microlensing timescale (blue line) caused by our Galactic halo PBHs to the stars in Euclid field view as a function of the PBH mass.}
\label{fig3}
\end{figure}

 In Fig. \ref{fig3} we plot the PBH  microlensing rate  and microlensing timescale towards the Euclid field of view as a function of the PHB mass.  We remind that the plot of the microlensing rate is given for $f_{PBH} = 1$, $u_T=6.54$ and  the microlensing timescale for $u_T=6.54$, the last one is independent of  $f_{PBH}$. If we see for the PBH mass $10^{-6}M_{\odot}$  (about the Earth-mass) the timeslace is $0.24~$days, included in the range  of timescale $[0.1,0.3]$ days found for $6$ microlensing events observed by OGLE towards the Galactic bulge (\cite{niikura2019microlensing}).\\
Considering a cadence of 20 minutes (\cite{laureijs2011euclid}) and no less than 5 points within $t_E$, we find that Euclid could detect PBHs down to $M=10^{-7} M_\odot$. \\
In the same way, we estimate also the smaller PBH mass detectable by the Subaru Hyper Suprime-Cam (HSC). Since its photometric accuracy is $\sim 1\%$  (\cite{huang2018characterization}), this leads  to $A_T \simeq 1.01$  and from eq. (\ref{A}) one can find $u_T \simeq 3.5$. The galactic coordinates of the HSC towards the M31 are: $(l, b) = (121.2^\circ, -21.6^\circ)$ (\cite{niikura2019constraints}).  Considering the $f_{PBH}=1$, we calculate the average optical depth and microlensing rate and then the microlensing timescale $t_E$.  For a cadence of 2 minutes and no less than 5 points within $t_E$, we find that HSC could detect down to $M=10^{-8} M_\odot$, a good match between theoretical calculations and observations.

\subsection{A mass function for PBHs}

Recently, Sureda et al. (\citeyear{sureda2020press}) derived a mass  distribution for PBHs, based on two different scenarios, named FCT (Fixed Conformal Time) and PPS (Primordial Power Spectrum). We consider their FCT mass function: 

\begin{equation}
\left(\frac{dn}{dM}\right)_{fct}=\frac{\rho_{DM}(a)}{\sqrt{2\pi}}\frac{n_s +3}{3}\frac{1}{M^2}\left(\frac{M}{M_\ast}\right)^{\frac{n_s +3}{6}} \times \exp \left[-\frac{1}{2}\left(\frac{M}{M_\ast}\right)^{\frac{n_s +3}{3}}\right].
\label{M31(B)}
\end{equation}
Here, $n_s=0.9649$ is the spectral index measured by the Planck collaboration, $\rho_{DM}(a)$ is the dark matter density for the scale factor today and $M_{\ast}=1.39\times 10^2 M_\odot$.

We  consider  the special  case of Euclid observations, 
in order to check the effects on the microlensing rate and, consequently, on the expected number of events, of a somehow  more realistic PBH mass distribution with respect to the monochromatic mass adopted in the previous sections of this paper.
For the  PBH mass range $[10^{-14},10^{2}]M_{\odot}$ we find, obviously, the same value for the optical depth as before: $<\tau>\simeq 1.29 \times 10^{-7}$ (indeed, the optical depth is a geometrical quantity). Concerning the microlensing rate, we find
$<\Gamma>\simeq 3.57 \times 10^{-7}$, which turns out to be a factor about 2.3 lower than the value ($8.12 \times 10^{-7}$) found previously for a fixed PBH mass of $1\, M_{\odot}$. 
 
\section{Conclusions}

Gravitational microlensing is, at present, the best method to
obtain valuable information about the mass function and the spatial distribution of PBHs, if they contribute to a high enough fraction of the Galactic halo dark matter. In this paper we aimed to investigate the gravitational microlensing traces of MW halo's PBHs towards four selected targets: the MW bulge, Large and Small Magellanic Clouds and the M31 galaxy. For PBHs we assume a standard halo model.  

We estimate  the optical depth and the microlensing rate towards these targets, assuming that the full halo dark matter is composed by PBHs, that is $f=1$ (but the generalization to any value of $f$ is straightforward), threshold impact parameter $u_T=1$ (the generalization to any value of $u_T$ is also straightforward) and assuming  a monochromatic  mass distribution for the PBHs.  The largest values  of both the optical depth and the microlensing rate are found for observations towards the M31 galaxy, due to its larger distance. SMC offers, as well, a rather large event rate.  

We calculate and discuss the microlensing timescale, which changes by many orders of magnitude, for the reason of the high mass range $[10^{-14},10^{2}]M_{\odot}$ of PBHs, taken into consideration. It varies from less than $1$ seconds up to $4.5$ years. Therefore, completely different observation strategies are necessary in order to detect PBHs in different bins of masses. For example, the Large Synoptic Survey Telescope (LSST), which is supposed to observe 
most of the visible sky every about $4$ days (\cite{sajadian2019predictions}), would not allow detecting microlensing events caused by the PHBs with masses smaller than  $0.1 M_\odot$ (no more than 5 points  within $t_E$). But, Subaru Hyper Suprime-Cam (HSC), with a cadence of $2$ minutes towards M31, could detect PBHs down to $M=10^{-8} M_\odot$  (\cite{niikura2019constraints}).

 As far as the Euclid observations  towards the Galactic bulge is concerned, and based on its features,  we find that  it could detect PBHs with mass down to $M=10^{-7} M_\odot$.  Below this mass limit, the expected number of points in the lightcurve is insufficient to perform a reliable analysis. For PBH masses around  this value we estimated a number of $340 x \sqrt{10^7} $ events per year, since the  microlensing rate varies as $1/\sqrt{M}$. \\
Our results are based on the assumption of a magnitude limit value of 24. In the case of observations shallower than that value, the number of bulge stars observed by Euclid gets  lower than $N_{ED}=6.45 \times 10^7$ and it would reduce sensibly the number of detectable microlensing events.\\
In the  case of Euclid, we also took into consideration an extended and more realistic  mass function  and find how the microlensing rate depends on the PBH mass function. This is certainly an important issue  to be considered during the planning of Euclid observations. 

Since a confusing element in the observed stars is the stellar intrisic variability, it needs a special attention. In fact, this can be avoided by carefully looking at different bands: gravitational microlensing is achromatic effect, while stellar variability appears different in different spectral bands. 

We have to remark that the detection of microlensing events toward M31 still remains mostly a theoretical possibility, for the reason of reduced capability of telescopes to resolve individual stars. Actually,  pixel-lensing surveys is required, which, in fact, do not alter the value of the calculated microlensing rate. However, in pixel lensing surveys the number of useful stars as targets is reduced since only the brightest stars can be observed as microlensed sources, thus reducing the number of expected microlensing events.

In our analysis we have considered neither the self-lensing events nor the contribution from PBHs in the halo of SMC, LMC or M31 galaxy. As far as for self-lensing events, they would mostly involve ordinary stars and not PBHs. They should be more numerous towards the inner part of those galaxies and have different time-duration (see, e.g., for the case of M31 the paper \cite{ingrosso2007new}). Of course, distinguishing self-lensing events from  halo events is always not easy and generally it has to be done on the basis of statistical considerations. Events due to PBHs in the halos around SMC, LMC and M31 have also been ignored.  Including these events would certainly increase the number of expected microlensing events. But a thorough analysis of this relevant issue is left to another, more comprehensive, paper.

  One important point to highlight here is how can we distinguish the different kinds of lensing object, that is among lenses  either in the target galaxy itself  or in the Galactic disk and a PBH in the Galactic halo. The appropriate way to solve this problem is to fully characterize a-posteriori each detected event by determining the event parameters (lens mass, Einstein time duration, lens  transverse velocity). This can be done by considering second-order microlensing  effects such as finite source effects, parallax, and astrometric effects along the line of sight   (\cite{hamolli2019free}). This is an objective for a next paper.

{\bf Acknowledgements} 

{\it FDP and AAN acknowledge the support by the INFN projects TAsP and Euclid. The authors also acknowledge the anonymous referee for his/her comments that significantly help improving the manuscript.}

\bibliographystyle{spr-mp-nameyear-cnd}
\bibliography{biblio-u1}

\end{document}